\documentclass[useAMS,usenatbib]{mn2e}
\usepackage{graphicx}
\usepackage{aas_macros}

\title[X-ray binaries in the Magellanic Bridge]{The Magellanic Bridge:  evidence for a population of X-ray binaries}
\author[V.A. McBride et al.]{V.A. McBride$^{1}$\thanks{E-mail:
vanessa@soton.ac.uk (VAM)}, A.J. Bird$^{1}$, M.J. Coe$^{1}$, L.J. Townsend$^{1}$, R.H.D. Corbet$^{2,3}$ and F.Haberl$^{4}$\\ 
$^{1}$School of Physics \& Astronomy, University of Southampton, SO17 1BJ, United Kingdom\\
$^{2}$University of Maryland, Baltimore County, MD, USA\\
$^{3}$CRESST/Mail Code 662, NASA Goddard Space Flight Center, Greenbelt, MD 20771, USA\\
$^{4}$Max-Planck-Institut f\"ur extraterrestrische Physik, Giessenbachstra{\ss}e, 85748 Garching, Germany}
\begin{document}

\date{Accepted 2009 December 7.  Received 2009 November 4; in original form 2009 September 24}

\pagerange{\pageref{firstpage}--\pageref{lastpage}} \pubyear{2002}

\maketitle

\label{firstpage}

\begin{abstract}
INTEGRAL observations of the Small Magellanic Cloud region have resulted in the serendipitous detection of two transient hard X-ray sources in the Magellanic Bridge.  In this paper we present the timing and spectral characteristics of these sources across the 2--100\,keV energy range, which, in conjunction with their optical counterparts, demonstrate that they are high mass X-ray binaries in the Magellanic Bridge.  Together with one previously known high mass X-ray binary system, and three candidates, these sources represent an emerging population of X-ray binaries in the Bridge, probably initiated by tidally induced star formation as a result of the gravitational interaction between the Large and Small Magellanic Clouds. 
\end{abstract}


\section{Introduction}
The Magellanic Clouds present a nearby, easily observable laboratory in which to study minor galaxy interactions.  In  this system, the Magellanic Bridge is thought to be a product of the tidal interaction between the Large and Small Magellanic Clouds (LMC \& SMC).  It contains both gas and stellar components, with the stellar population first discovered by \citet{IrwinKunkelDemers1985} and comprising only a young ($<300$\,Myr) stellar component \citep{Harris2007}, which is thought to have formed in situ.  One identified massive X-ray binary has been detected in the western bridge \citep{KahabkaHilker2005}.  Given the tidal disruption in this region, a number of star formation tracers are expected in this active region.  The focus of this work is to identify high mass X-ray binaries as tracers of star formation in the Magellanic Bridge.

Multiwavelength studies of the SMC have shown that it contains a large number of X-ray binary pulsars -- all but one of them in Be/X-ray binary systems \citep{CoeCorbetMcGowan2009,CoeEdgeGalache2005}.  This preponderance of young systems is most likely explained by tidally triggered star formation precipitated by the most recent close approach of the SMC and LMC \citep{GardinerNoguchi1996}.  However, the most recent close approach of the SMC and LMC was $\sim200$ Myr ago -- much longer than the evolutionary timescale of Be/X-ray binaries, which implies that either there has been a significant delay between the encounter of the SMC and LMC and the onset of star formation, or that subsequent waves of star formation have given rise to these Be/X-ray binaries \citep{HarrisZaritsky2004}.  These objects, being tracers of star formation \citep{GrimmGilfanovSunyaev2003}, give direct insights into the star formation history of their host galaxies.

The LMC, by comparison, has a very different population at high energies:  whereas the SMC population comprises almost exclusively Be/X-ray binaries, the population of the LMC has representatives from all members of the X-ray binary classes, including black hole systems, low mass X-ray binaries, and Be and supergiant high mass systems.  With the mass of the LMC being $\sim10\times$ that of the SMC, tidal interactions between the galaxies would have a much greater effect on the SMC, and this may be reflected by the differing stellar populations. 

The high energy population of the Magellanic Bridge is not well known, but extrapolation from optical wavelengths indicates that there should be many young stellar systems towards the western bridge \citep{Harris2007}, resembling the population in the SMC.  The work presented here is the first systematic study of this region at hard X-ray energies.  The rest of the paper is structured as follows:  section 2 presents the observations and data analysis.  In section 3, the new transient objects and their characteristics are presented, while other candidate high energy sources are discussed in section 4.  The discussion and conclusions are presented in sections 5 and 6 respectively.

\begin{figure*}
 \includegraphics[width=0.7\textwidth]{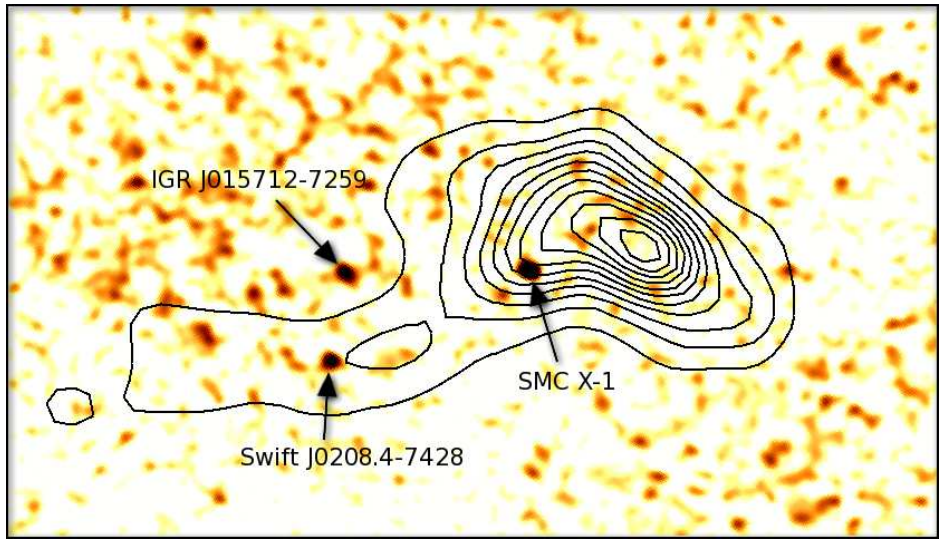}
 \caption{IBIS 15--35\,keV mosaic of the SMC and Magellanic Bridge from revolutions 753 -- 756.  The contours show the H\,I distribution in the SMC and Bridge, and are from \citet{PutmanStaveley-SmithFreeman2003}. }
 \label{fig:mosaic}
\end{figure*}

\section{OBSERVATIONS \& ANALYSIS}
The IBIS telescope \citep{UbertiniLebrunDiCocco2003} aboard \textit{INTEGRAL}, which is optimised for an energy range of 15--200\,keV and has a field of view of $30^\circ\times30^\circ$, is uniquely suited to observing large sky areas for point sources.  As part of a key programme monitoring campaign on the SMC and 47 Tuc, \textit{INTEGRAL} observed the SMC and Magellanic Bridge for approximately 90\,ks per satellite revolution ($\sim3$ days) from 2008 November 11 to 2009 June 25.   The exposure on the Magellanic Bridge (Table~\ref{tab:expo}) is, in general, much smaller than the 90\,ks observation time, both due to the fact that the Bridge is a few degrees from the main pointing position, and that the figures for exposure time have been corrected for instrument dead time.

Individual pointings (science windows) were processed using the \textit{INTEGRAL} Offline Science Analysis v.7.0 (OSA, \citealt{GoldwurmDavidFoschini2003}) and were mosaicked using the weighted mean of the flux in the 3--10\,keV (JEM-X) and 15--35\,keV (IBIS) energy ranges.  Proprietary software was used to mosaic the observations from successive revolutions to improve the exposure and thereby the sensitivity to faint sources.  Lightcurves in these energy bands were generated on science window ($\sim$2000\,s) and revolution time-scales. The IBIS energy band was chosen to maximise the detection significance of \hbox{SMC~X-1}, and hence other SMC accreting X-ray pulsars, which have similar spectral shapes to SMC~X-1 in this energy range.  An IBIS mosaic of data from revolutions 752--756 in the 15--35\,keV is shown in Fig.~\ref{fig:mosaic}.

%%%%%%%%%%%%%%%%%%%%%%%%%%%%%%%%
%Crab flux is 9.4E-9 erg/cm2/s and 248 counts/s in the 15-35 keV band
%%%%%%%%%%%%%%%%%%%%%%%%%%%%%%%%%%

\begin{table}
\caption{INTEGRAL observation log.}
\label{tab:expo}
 \begin{tabular}{llll}
  \hline
   Rev Number & MJD$_{\rm start}$ & Exposure (ks) &  \\
  \hline
% Exposures (from survey_mosaic7)
 745 & 54788.7 & 37.7 & \\
 746 & 54791.8 & 40.2 & \\
 747 & 54794.8 & 37.0 & \\
 748 & 54797.7 & 41.1 & \\
 749 & 54800.8 & 39.7 & \\
 750 & 54803.7 & 46.0 & \\
 751 & 54806.7 & 49.2 & \\
 752 & 54809.7 & 51.9 & \\
 753 & 54812.7 & 50.0 & \\
 754 & 54815.7 & 39.2 & \\
 755 & 54818.7 & 50.6 & \\
 756 & 54821.7 & 49.5& \\
 796 & 54941.4 & 69.8 & \\
 797 & 54944.4 & 55.6 & \\
 812 & 54989.2 & 49.5 & \\
 813 & 54992.2 & 40.2 & \\
 814 & 54995.2 & 49.3 & \\
 815 & 54998.2 & 48.7 & \\
 816 & 55001.2 & 52.9 & \\
 817 & 55004.2 & 55.4 & \\
 818 & 55007.2 & 31.5 & \\
 \hline
 \end{tabular}
\end{table}

\section{New sources}

\subsection{IGR~J015712$-$7259}
\begin{figure}
 \centering
 \includegraphics[width=84mm]{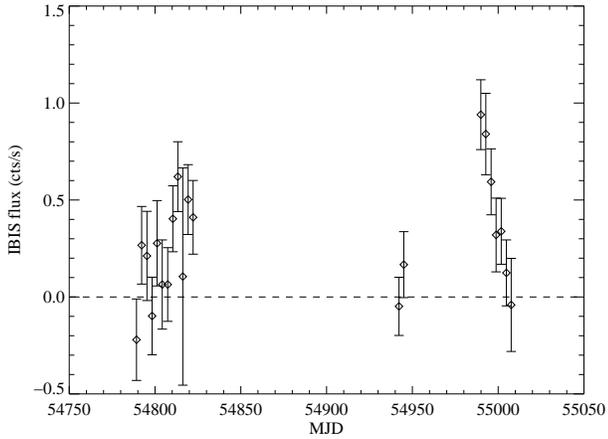}
 \caption{IBIS lightcurve of IGR J015712$-$7259 in the 15--35\,keV energy band.}
 \label{fig:igrlc}
\end{figure}

\begin{figure}
 \centering
\fbox{ \includegraphics[width=84mm]{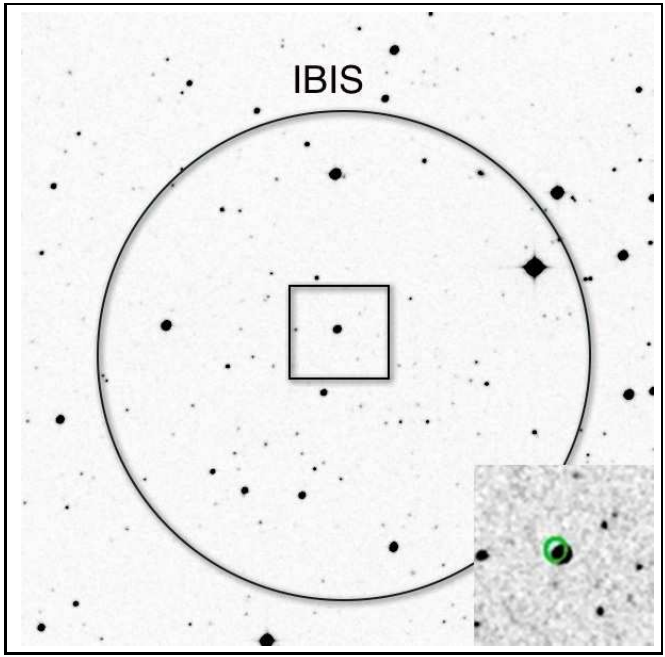}}
 \caption{DSS II R band image of a 10$^\prime\times10^\prime$ field around the source IGR~J015712$-$7259, showing the IBIS 90\% error circle and with the inset showing an expanded view (indicated by the square) of the counterpart and the Swift XRT 90\% error circle.  North is up and east is to the left.}
 \label{fig:igrfinder}
\end{figure}

A new source, IGR~J015712$-$7259 \citep{CoeMcBrideBird2008} was discovered in revolution 752.  It showed transient behaviour with a maximum significance of 7.7$\sigma$ from revolutions 812 to 814 in IBIS, corresponding to an average flux of $3.3\pm0.3\times10^{-11}$\,erg\,cm$^{-2}$\,s$^{-1}$ ($1.4\times10^{37}$\,erg\,s$^{-1}$ at 60\,kpc).%0.814+/-0.105
  The source was detected in JEM-X at an average flux of $1.6\pm0.5\times10^{-12}$\,erg\,cm$^{-2}$\,s$^{-1}$ in 3--10\, keV.  The IBIS lightcurve is shown in Fig.~\ref{fig:igrlc}.  Timely follow-up with Swift/XRT on 2008 December 20 (MJD 54820) allowed a precise determination of the source position: 1$^{\rm h}$57$^{\rm m}$16$^{\rm s}$, $-72^\circ$58$^\prime$32.8$^{\prime\prime}$ (J2000.0) 
with a 90\% error circle of radius 3.8$^{\prime\prime}$.   This identifies the source with a star (USNO-B1 0170-0064697, \citealt{MonetLevineCanzian2003}) having B and R magnitudes of 15.48 and 15.51 respectively (see Fig.~\ref{fig:igrfinder}.)  A total of $\sim$300 counts extracted from a region around the Swift/XRT source showed pulsations at $\sim11$\,s. 

\begin{figure}
 \centering
\includegraphics[width=84mm]{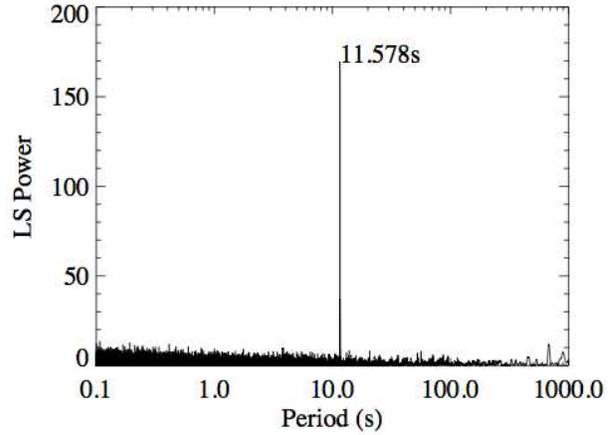} 
 \caption{Periodogram of the \textit{RXTE} observation of IGR~J015712$-$7259 on MJD 54824.}
 \label{fig:SXP11.6}
\end{figure}

\begin{figure}
\centering
\includegraphics[width=84mm]{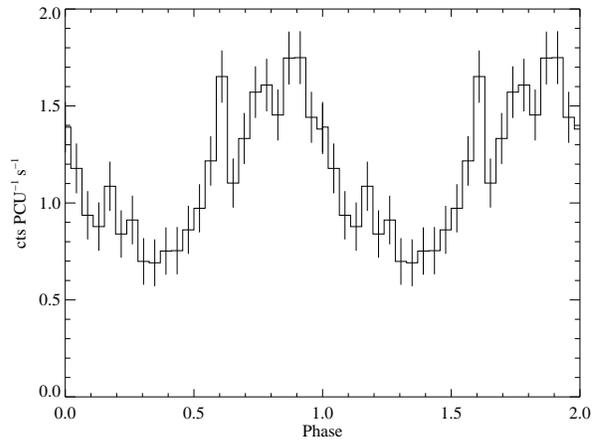}
\caption{RXTE 3--10\,keV lightcurve of IGR J015712$-$7259 folded at the pulse period of 11.578\,s}
\label{fig:p_profile}
\end{figure}

\begin{figure}
 \centering
 \includegraphics[width=84mm]{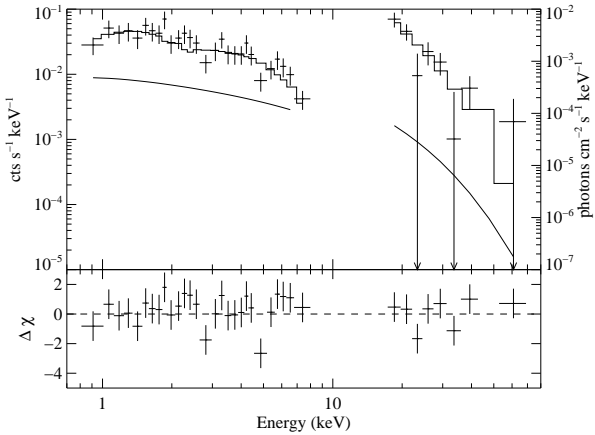}
 \caption{Combined Swift/XRT and IBIS spectrum of IGR~J015712$-$7259, showing the residuals as $\Delta\chi$ in the lower panel.  The data are plotted as crosses, the model fit as the histogram, while the smooth curve shows the unfolded spectrum as plotted on the secondary Y-axis.}
 \label{fig:igrspec}
\end{figure}

A 15\,ks RXTE observation on 2008 December 24 %JD 2454825.09250452 
revealed that the source was pulsating with a period of 11.57809$\pm$0.00002\,s
(see Fig.~\ref{fig:SXP11.6}).  The hard X-ray behaviour is illustrated in the IBIS lightcurve (Fig.~\ref{fig:igrlc}), the RXTE folded lightcurve (Fig.~\ref{fig:p_profile}) and the broadband spectrum of the source around MJD 54824 (Fig.~\ref{fig:igrspec}).  The combined Swift/XRT and IBIS spectrum can be adequately fit ($\chi^2_\nu=1.05$ for 33 d.o.f.) with an absorbed, exponentially cut-off power law and a free constant factor between the two instruments to account for the fact that the soft and hard spectra were not observed simultaneously.  The constant factor is $\sim0.3$, while the photon index is $0.4\pm0.2$, and the folding energy is $8^{+5}_{-3}$\, keV.  The absorption has been fixed to the neutral density of hydrogen along the line of sight to the SMC:  $6\times10^{20}$\,cm$^{-2}$.  The photon index is similar to those determined for X-ray pulsars in the wing of the SMC \citep{McGowanCoeSchurch2007}. The transient nature of the source, combined with the detected pulsations, hard spectrum and the optical magnitudes of the counterpart are strongly suggestive of a Be/X-ray binary.

\subsection{SWIFT~J0208.4-7428}
Another source was discovered during revolution 756 (MJD 54821.7) in the IBIS map at a position of $02^{\rm h}07^{\rm m}09^{\rm s}$ $-74^\circ28^\prime07^{\prime\prime}$with a 3.6$^\prime$ error circle. It reached a maximum significance of 7.1$\sigma$ during revolutions 753--756, with a corresponding average flux of {2.5$\times10^{-11}$\,erg\,cm$^{-2}$\,s$^{-1}$ %0.618 counts/s
 in the 15-35\,keV band ($\sim1\times10^{37}$\,erg\,s$^{-1}$ at 60\,kpc).  The source was weakly detected with JEM-X, at a flux of 1.5$\pm0.6\times10^{-12}$\,erg\,cm$^{-2}$\,s$^{-1}$. %0.0103 counts/s. 
The lightcurve in the 15--35\,keV band in Fig.~\ref{fig:swift0208lc} clearly shows the flux rising through the observation sequence.

Archival data from Swift/XRT showed that a source in the Magellanic Bridge was observed with Swift on MJDs 54764.5 and 54809.5.  Analysis of the XRT data shows that no source is present within the field of view during the observation on MJD 54764.5, but during MJD 54809, a source is detected at $02^{\rm h}06^{\rm m} 45.7^{\rm s}$, $-74^\circ27^\prime 46.3^{\prime\prime}$  J2000.0 with a 90\% error circle of 4$^{\prime\prime}$.  This source falls well within the 90\% error circle of the INTEGRAL source, and can be clearly distinguished from RX~J0209.6$-$7427 (the only previously known HMXB in the Magellanic Bridge, see Fig.~\ref{fig:swift0208_finder}).%is 1.6 arcmin away
These contemporaneous observations (see lightcurve in Fig.~\ref{fig:swift0208lc}) provide strong evidence for another new X-ray source in the Magellanic Bridge. Only one object falls within the Swift/XRT error circle and can be identified with a $B=14.41$ and $V=14.75$ star \citep{DemersIrwin1991}, for which the $B-V$ colour and magnitudes are consistent with an early-type dwarf at the distance of the SMC. 

\begin{figure}
 \centering
 \includegraphics[width=84mm]{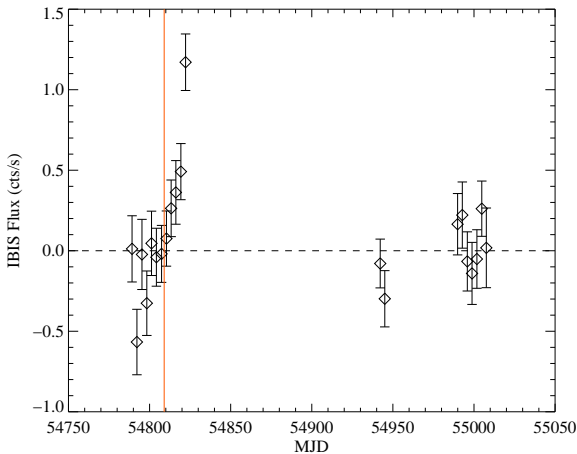}
 \caption{IBIS lightcurve of SWIFT~J0208.4$-$7428 (red line shows Swift/XRT detection).}
 \label{fig:swift0208lc}
\end{figure}

\begin{figure}
 \centering
\fbox{ \includegraphics[width=84mm]{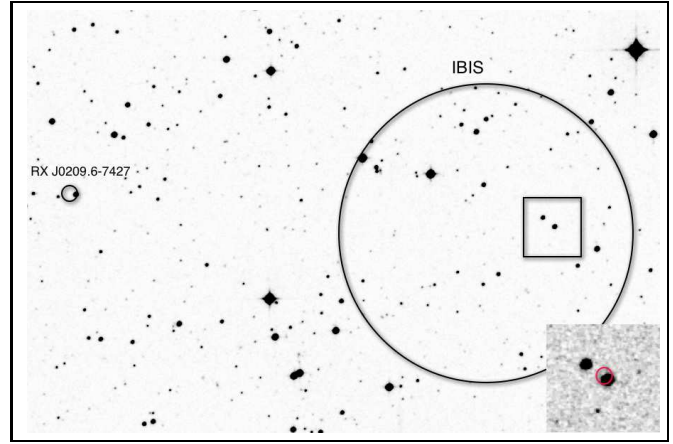}}
 \caption{DSS II R-band image of a $10^\prime\times15^\prime$ field surrounding Swift~J0208.4$-$7428, showing the $3.6^\prime$ radius 90\% IBIS error circle, the nearby known HMXB, RX~J0209.6$-$7427, and with the inset showing an expanded view of the square region along with the Swift XRT error circle.  North is up and east is to the left.}
 \label{fig:swift0208_finder}
\end{figure}

\section{Candidate sources}
IBIS 15--35\,keV and JEM-X 3--10\,keV maps on time-scales varying between single revolutions up to five consecutive revolutions  were searched for excesses which could indicate potential new sources in the Magellanic Bridge. A list of candidates is supplied in Table~\ref{Tab:cands}.  None of these candidate sources correlate with known ROSAT sources \citep{VogesAschenbachBoller1999,VogesAschenbachBoller2000}, but it is worth bearing in mind that the ROSAT area coverage in this part of the sky is not complete.  With the large error circles on these objects, it is not yet possible to constrain the nature of these sources, and, although  they may be background AGN, their transient nature makes it more likely that they are binary systems in the Magellanic Bridge.

\begin{table*}
\caption{Candidate sources in the Magellanic Bridge. Columns 5 and 6 indicate the revolutions that were mosaicked for each candidate detection.  For IBIS detections the fluxes are given in the 15--35\,keV range, while JEM-X fluxes are quoted in the 3--10\,keV range.}
\begin{tabular}{l l l l l l l}
\hline
Name & RA (J2000.0) & Dec (J2000.0) & Error & IBIS & JEM-X & Flux (Error)\\
 & \emph{h\,m\,s} & \emph{d\,m\,s} & 90\% & Revs & Revs & erg\,cm$^{-2}$\,s$^{-1}$ \\
\hline
IGR J02048$-$7315 & 02 04 49 &-73 15 27 & 1.8$^\prime$ & &754--756 & $1.7(0.7)\times10^{-12}$ \\
IGR J02220$-$7558 & 02 22 01 & -75 57 59 & 1.8$^\prime$ & & 755--756 & $3.5(1.1)\times10^{-12}$\\ %(J11)
IGR J03144$-$7404 & 03 14 23  &-74 04 23& 4.9$^\prime$ & 746--747&  &  $4.4(0.9)\times10^{-11}$ \\%2 arcmin from 10th mag star (C1)\\
\hline
\end{tabular}
\label{Tab:cands}
\end{table*}

\section{Discussion}
Before this work, only a single massive X-ray binary system was known in the Magellanic Bridge (RX~J0209.6$-$7427,  \citealt{KahabkaHilker2005}).  So far, all X-ray sources discovered and identified in the Magellanic Bridge have been transient, with no persistent sources detected down to a luminosity of 2.5$\times10^{36}$\,erg\,s$^{-1}$ (15--35\,keV).  In this respect the population of the Magellanic Bridge seems to resemble that of the SMC more closely than it does the LMC. The Magellanic Bridge contains a significant population of young blue stars \citep{DemersIrwin1991,Harris2007}.  \citet{Harris2007} showed that the distribution of blue stars is most dense towards the SMC side and that there appear to be very few old, red giant branch stars in the fields studied.  As most HMXBs in the SMC are found in regions populated by young stars \citep{YokogawaImanishiTsujimoto2003}, this seems to indicate that the trend will be to find more HMXBs on the SMC side of the Bridge, where the stellar population is younger.  However, in our work, the exposure is by no means uniform across the Magellanic Bridge (see Fig.~\ref{fig:expo}) and this introduces a selection effect whereby new HMXBs are selectively found where the exposure is highest, i.e. towards the SMC side of the Bridge.  Thus, we cannot yet be confident that the population of X-ray binaries so far uncovered in the Bridge is indicative of the distribution in the Bridge. 

The prevalence of a young stellar population, and lack of an older one, suggests that the stellar population of the Magellanic Bridge formed in-situ, rather than being tidally extracted from the Large and Small Magellanic Clouds.  In addition to the HMXBs discovered here, the region also shows strong evidence for recent star formation with $\sim100$ OB associations identified in the western Magellanic Bridge \citep{BicaSchmitt1995} as well as molecular clouds \citep{MizunoMillerMaeda2006}, which act as fuel for ongoing star formation.

\begin{figure}
\centering
\fbox{\includegraphics[width=84mm]{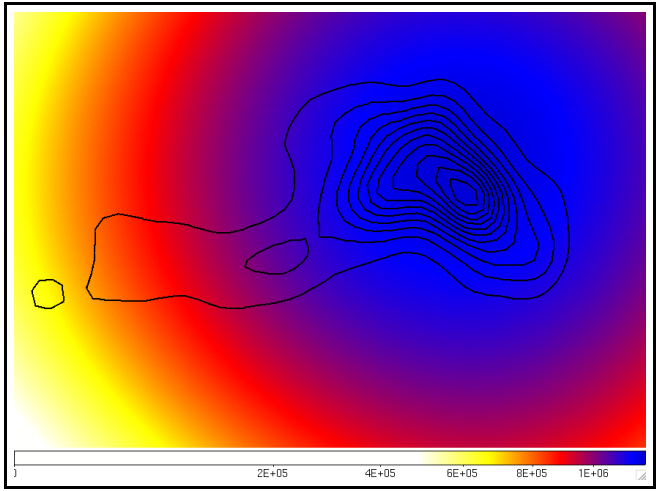}}
\caption{IBIS exposure map for the sum of all observations presented in this paper.  The exposure on the eastern edge of the bridge is almost half of that on the western edge (SMC side).  The image size is $10^\circ\times15^\circ$, and the contours once again show the H\,I distribution throughout the SMC and Magellanic Bridge.}
\label{fig:expo}
\end{figure}

It is very unlikely that the HMXBs so far discovered in the Magellanic Bridge could have been ejected from the SMC through supernovae kicks.  Not only would they require systemic velocities in excess of 200\,km\,s$^{-1}$ (where most SMC HMXBs have estimated systemic velocities of $\sim30$\,km\,s$^{-1}$, \citealt{Coe2005}), but we would expect to find a distribution of HMXBs around the SMC rather than preferentially concentrated in the Bridge.

To estimate what fraction of a potential HMXB population in the Magellanic Bridge has been observed to date, we use an estimate of 1 Be+NS binary system per square degree, as obtained by \citet{KahabkaHilker2005}.  This estimate is based on the observed stellar density of stars $>2.5M_\odot$ in the Magellanic Bridge \citep{GardinerHatzidimitiou1992}, corrected through the initial mass function to give the number of B stars per square degree.  Through estimates of the Be/B fraction and an assumption of how many Be stars have a neutron star companion, these authors estimate the number of Be+NS systems as between 0.68 and 0.96 per square degree, dependent on the shape of the IMF.  Given that the Magellanic Bridge spans an area of $\sim6$ square degrees on the sky, we may expect up to 6 Be+NS systems in the Magellanic Bridge, which is roughly consistent with the number of sources and potential candidates reported up until now, including this work.  At this early stage, it is premature to attempt quantitative estimates of an upper limit to the potential number of HMXBs in the Magellanic Bridge.  A prediction based on star formation rate (e.g. \citealt{GrimmGilfanovSunyaev2003}) is prone to significant uncertainties in both the X-ray luminosity function of this potential population, and in estimates of the star formation rate.   However, the X-ray active population in the SMC during the same set of observations, is very similar in number to that of the Magellanic Bridge.  In the SMC, only four out of the $\sim60$ HMXBs were X-ray active during these observations, so by observational comparison with the SMC, we may expect a population significantly larger than 6 Be/X-ray binaries in the Bridge.

\section{Conclusions}
In this paper, we presented two new hard X-ray sources in the Magellanic Bridge identified through wide-field, hard X-ray imaging.  We described their timing and spectral characteristics and identified them as likely high mass X-ray binaries located at the distance to the Magellanic Bridge.  This puts the number of high mass X-ray binaries in the Magellanic Bridge at three, with a further three candidate sources to add to the emerging population.  Optical observations of this region show a large number of young stars, especially towards the western edge, consistent with a picture of in-situ star formation as result of tidal interaction between the SMC and LMC.  The short-lived high mass X-ray binaries presented here echo this scenario for star formation in the Bridge.

\section*{Acknowledgements}
Based on observations with \emph{INTEGRAL}, an ESA project with instruments and science data centre funded by ESA members states (especially the PI countries: Denmark, France, Germany, Italy, Switzerland, Spain), Czech Republic and Poland, and with the participation of Russia and the USA.  LJT wishes to the University of Southampton, whose support has made this research possible.
\bibliographystyle{mn2e}

\label{lastpage}

\end{document}